\documentclass[reprint,aps,prl,superscriptaddress,english,nolongbibliography]{revtex4-1}

\usepackage{graphicx,dcolumn,braket}
\usepackage[table,xcdraw,dvipsnames]{xcolor}
\usepackage[colorlinks=true,citecolor=blue,urlcolor=blue,linkcolor=blue]{hyperref}
\usepackage[caption=false,subrefformat=parens,labelformat=empty]{subfig}
\usepackage{amsmath,amssymb,bm,mathrsfs,amsfonts}
\usepackage{verbatim}

\begin{document}
\title{Renormalization Group Analysis of the Anderson Model on Random Regular Graphs}

\author{Carlo Vanoni}
\email{cvanoni@sissa.it}
\affiliation{SISSA -- International School for Advanced Studies, via Bonomea 265, 34136, Trieste, Italy}
\affiliation{INFN Sezione di Trieste, Via Valerio 2, 34127 Trieste, Italy}
\author{Boris L.\ Altshuler}
\affiliation{Physics Department, Columbia University, 538 West 120th Street, New York, New York 10027, USA}
\author{Vladimir E.\ Kravtsov}
\affiliation{ICTP, Strada Costiera 11, 34151, Trieste, Italy}
\author{Antonello Scardicchio}
\affiliation{ICTP, Strada Costiera 11, 34151, Trieste, Italy}
\affiliation{INFN Sezione di Trieste, Via Valerio 2, 34127 Trieste, Italy}

\begin{abstract}
    We present a renormalization group analysis of the problem of Anderson localization on a Random Regular Graph (RRG) which generalizes the renormalization group of Abrahams, Anderson, Licciardello, and Ramakrishnan to infinite-dimensional graphs. The renormalization group equations necessarily involve two parameters (one being the changing connectivity of sub-trees), but we show that the one-parameter scaling hypothesis is recovered for sufficiently large system sizes for both eigenstates and spectrum observables. 
    We also explain the non-monotonic behavior of dynamical and spectral quantities as a function of the system size for values of disorder close to the transition, by identifying two terms in the beta function of the running fractal dimension of different signs and functional dependence. Our theory provides a simple and coherent explanation for the unusual scaling behavior observed in numerical data of the Anderson model on RRG and of Many-Body Localization.
\end{abstract}

\maketitle

Recent works on many-body localization (MBL) \cite{Basko06,Pal10,Serbyn13b,ros2015integrals,Nandkishore15,Imbrie17,Abanin2019colloquium} have challenged our understanding of thermalization in quantum, disordered systems. Hamiltonian systems under strong disorder display breakdown of ergodicity and show absence of transport or otherwise extremely slow, sub-diffusive dynamics \cite{vznidarivc2016diffusive}. When turning on the interaction on a system that is localized in the one-electron approximation, the perturbation theory presented in Ref.~\cite{Basko06} has many features that resemble the spreading of a quantum particle on an infinite-dimensional graph \cite{altshuler1997quasiparticle}, which can locally be approximated by a tree. It is therefore not surprising that the problem of the Anderson model~\cite{Anderson1958absence} on tree-like, infinite-dimensional structures has seen a revival~\cite{de2014anderson,parisi2019anderson,biroli2017delocalized,tikhonov2021AndersonMBL,Lemariè2022critical,baroni2023corrections,Mata2017Scaling,Mata2020two,tikhonov2016fractality,arenz2023wegner,Altshuler2016Nonergodic,Pino2020Scaling,zirnbauer2023wegner} as a consequence of the works on MBL. A remarkable milestone that determined development of the field of Anderson localization for decades was a scaling theory~\cite{abrahams1979scaling} by Abrahams, Anderson, Licciardello and Ramakrishnan, also known as the ‘gang of four' work. There, the ideas of real space renormalization group (RG) were successfully applied to the localization problem and one-parameter scaling was suggested, which is the most celebrated concept in the theory of Anderson localization in finite-dimensional systems.

Random graphs \cite{bollobas1998random}, in which edges are added randomly to a given set of vertices to satisfy a predefined set of conditions, are commonly used in statistical physics to define mean field models. This happens because they locally look like Cayley trees/Bethe lattices, and therefore recursion equations can be written for their statistical properties (see for example \cite{abou1973selfconsistent,aizenman2006absolutely} for the Anderson localization problem, \cite{thouless1977solution,chayes1986mean,mezard1987spin,mezard2001bethe} for the spin glass problem). Regular Random Graphs (RRGs) are  random graphs in which every vertex has the same coordination or connectivity $\mathcal{D}=K_0+1$ (see Fig.~\ref{fig:tree}). $K_0$ is usually called the {\it branching} number but we will refer to it as connectivity as well, neglecting the difference of 1 between the two definitions.

The geometry of RRGs, being expander graphs~\cite{kowalski2019introduction} of formally infinite dimension, behaves peculiarly under the block transformation of the renormalization group. Unlike a $d$-dimensional cube~\cite{Tarquini2017critical}, which is always connected to $2d$ other cubes, irrespective of their size, when we divide an RRG of connectivity $K_0$ in blocks of linear dimension $L$ (much smaller than its diameter), such blocks will have connectivity $K_0^L$ (see Fig.~\ref{fig:tree}). Connectivity is an important parameter in the Anderson model since, to a first approximation, localization is achieved when the disorder strength $W$ measured in units of hopping rate is much larger than $K_{0}$. Therefore, under block decimation or composition (to follow Ref.~\cite{abrahams1979scaling} and subsequent works \cite{lee1979real}) one needs to keep track of the ever-growing connectivity.

This {\it additional parameter} in the RG equations on expander graphs makes a big difference in terms of phenomenology, explaining what happens in the $d\to\infty$ limit of the equations in \cite{abrahams1979scaling}, and why, in this limit, the Anderson transition is akin to the Berezinski-Kosterlitz-Thouless transition \cite{1971JETP...32..493B,1972JETP...34..610B,1973JPhC....6.1181K}. Similar phenomenological RG equations have been conjectured to underlie the MBL transition \cite{PhysRevX.5.031032,PhysRevX.5.031033,PhysRevB.93.224201,Dumitrescu2019KT}, but this time the connection came from an analogy with the {\it strong disorder} Ma-Dasgupta-Hu-Fisher RG equations \cite{ma1979random,dasgupta1980low,fisher1994random,huse2023strongrandomness}. It is not surprising that the ``gang of four" RG equations \cite{abrahams1979scaling} should be modified and become similar to those of a many-body problem, as Cayley trees/Bethe lattices have been recognized as proxies of quantum dots \cite{altshuler1997quasiparticle,sivan1994quasi} and spin chains \cite{Basko06,ros2015integrals,de2014anderson,tikhonov2021AndersonMBL}.

In this work we show how, by considering the renormalization group approach, it is possible to interpret in a novel way the finite-size scaling of eigenstates observables and spectral indicators. Where the RG $\beta$-function cannot be completely fixed by theoretical arguments we rely on the state-of-the-art numerical results, presented in Ref.~\cite{sierant2023universality}, to extract the missing information we need.  We find that for $W<W_{c}$ the two-parameter scaling, present at smaller system sizes, reduces to a one-parameter scaling for sufficiently large sizes.  However, in contrast with the usual phase transitions in statistical models, the insulator phase for $W>W_{c}$ and the critical behavior at $W\lesssim W_c$ are beyond the single-parameter scaling curve. This is due to the fact, as we will make clearer in the following, that the insulating phase is a line of critical points and that critical phase is at the terminal point of such line and cannot be distinguished by it. In this case, it is necessary to split the $\beta$-function of any observable in two terms.  One of them, $\beta_0$, does not contain the system size (or the connectivity $K$) explicitly, and it governs the one-parameter scaling at large system sizes. The remainder, called $\beta_1$, will instead depend explicitly on $K$, and it does describe the two-parameter regime that becomes dominant close to $W_c$ (see also Refs.~\cite{arenz2023wegner,zirnbauer2023wegner}).

A detailed analysis of the numerical results in Ref.~\cite{sierant2023universality} allows us to accurately describe the $\beta$-function in large neighborhood of the ergodic fixed point, while the behavior close to insulating and critical region is not accessible by the available numerics. We, therefore, present some possible scenarios for the functional form of the $\beta$-function close to the critical point, explaining the consequences of each scenario for the critical exponents of the transition.
\section*{Renormalization Group Equations}
We consider the Anderson model on a RRG of connectivity $K_0$ (i.e. fixed vertex degree $\mathcal{D}=K_0+1$), defined by the Hamiltonian 
\begin{equation}
    H = - \sum_{\langle i,j \rangle} \left( \ket{i}\bra{j} + \ket{j}\bra{i} \right) + \sum_i \epsilon_i \ket{i} \bra{i},
\end{equation}
where $\epsilon_i$ are independent and identically distributed random variables sampled according to the box distribution $g(\epsilon) = \theta(|\epsilon|-W/2)/W$. Since in an RRG each vertex has a fixed connectivity, it is locally a Cayley tree, while on large scales loops will become important to ensure the regularity of the graph. If $\mathcal{N}$ is the number of vertices of the graph, it is possible to introduce a length scale $L = \log_{K_0} \mathcal{N}$, representing the diameter of the graph, i.e. the maximal length of the shortest paths connecting two nodes.

\begin{figure}
    \centering
    \includegraphics[width=0.9\columnwidth]{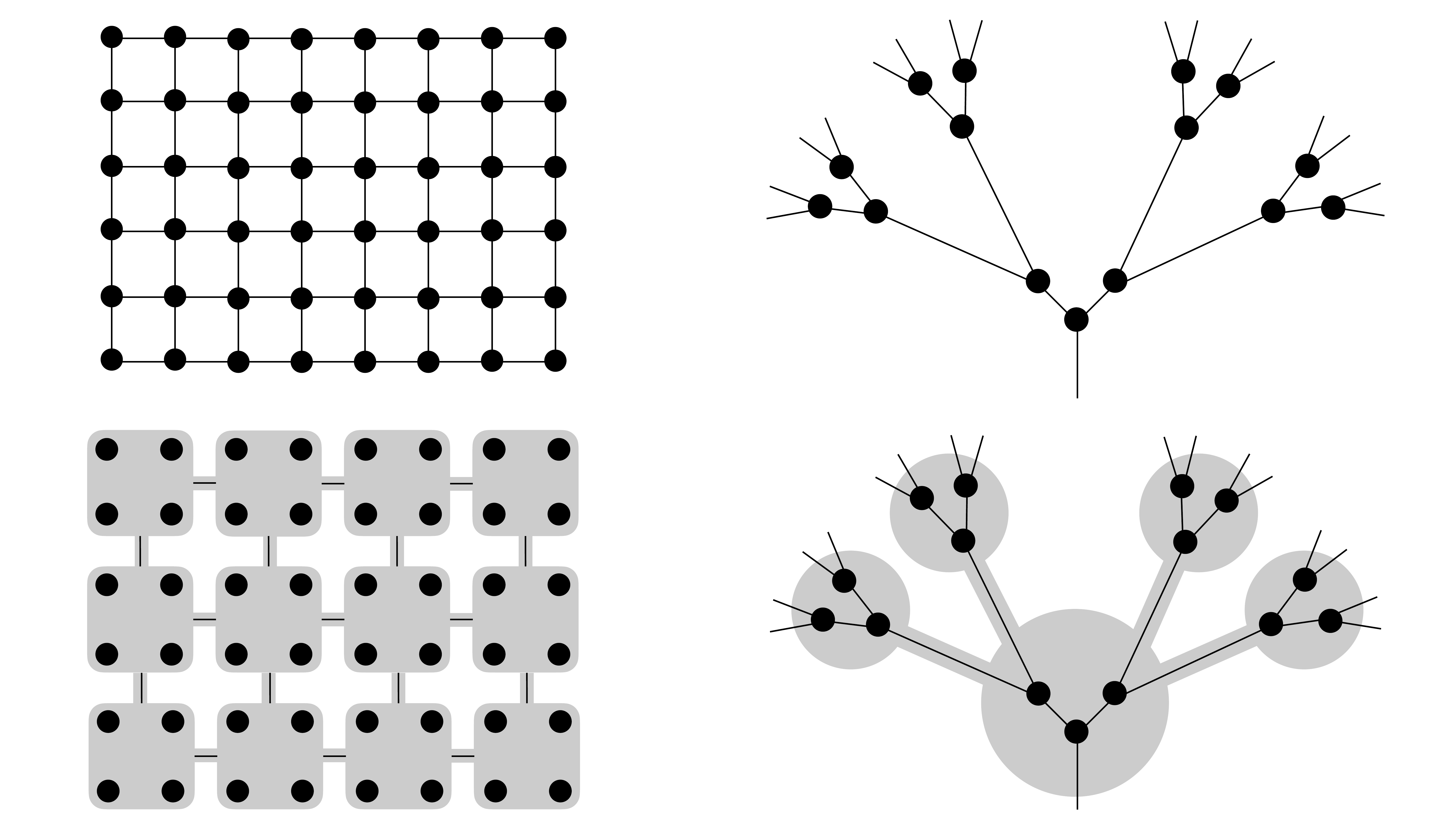}
    \caption{Grouping of sites under renormalization group. ({\it Left}) In finite dimensions, the connectivity of the blocks does not change, under RG block transformation. ({\it Right}) The RG transformation on a tree instead changes the connectivity of a block. One goes from $K_0$ (in the drawing $K_0=2$) to $K_0^L$ when sites at distance $L$ are grouped in the same effective node.}
    \label{fig:tree}
\end{figure}

Starting from a tree with connectivity $K_0$ (see Fig.~\ref{fig:tree}, where $K_0=2$) and proceeding in the spirit of the Kadanoff decimation procedure, we group subtrees of increasing depth creating new ``effective" nodes.
At step $L$, due to the Cayley tree geometry, the new node will have a larger coordination number $\mathcal{D} = K(L)+1$, which coincides with the number of nodes at distance $L$ in the original bare graph. This is the main difference with the situation in finite dimensions $d$, where the geometrical datum of the connectivity is independent of the renormalization scale $L$.
According to this blocking procedure, the equation for the connectivity $K(L)$ at step $L$ is simply  
\begin{equation}\label{Eq-K}
    \frac{d K}{d \ln L}=K \,\ln K,  
\end{equation}
This equation has the desired solution $K(L)=K_0^L$ which reflects the geometry of a local tree. This equation represents the geometric datum of the RRG at scale $L$ and we consider it now decoupled from the physical datum describing the structure of the eigenfunctions, the spectrum, or transport properties (like the conductance $g$ of Ref.~\cite{abrahams1979scaling}) at the same scale. We will content ourselves with this approach, although it is possible that, in the future, on the way towards an analytic solution, one might need to write directly coupled differential equations for effective geometric and physical quantities. Our simplification turns out to be sufficient in the metallic phase which we are mostly concerned with, so we will use it in the rest of the paper. 

As a physically meaningful second parameter, Ref.~\cite{abrahams1979scaling} would use the dimensionless conductance $g=L^{d-2}\sigma (\hbar/e^{2})$, where $\sigma$ is the sample conductivity. We, however, will consider eigenstates and eigenvalue (spectral) properties, for two reasons. On one hand, they are more easily accessed in modern numerical calculations, on the other they represent intrinsic properties of the unitary dynamics of our system, while the conductance is affected by the form of the coupling to the leads {\it etc.}. Eigenstate and spectral properties must be qualitatively (and quantitatively) determined by the expected number of resonant sites at a fixed energy $E$ (for example let us take the center of the band $E=0$) within a distance $L$. We can formally define the quantity $\psi$ as follows. By denoting a normalized wavefunction at site $i$ as $\varphi(i)$, following Refs.~\cite{support_set,kravtsov2018non}, we define the {\it support set} $S_{\varepsilon}$ satisfying the relation $\sum_{i \in S_{\varepsilon}} |\varphi(i)|^2 = 1 - \varepsilon$, where the $\varphi$'s entering the sum are the largest ones in modulus. The dimension of the set $S_{\varepsilon}$ is given by $K_S(\varepsilon) = \sum_{i \in S_{\varepsilon}} 1$~\cite{kutlin2024investigating}. $S_{\varepsilon}$ does not contain explicitly a length (except the system size), so we need to better describe its structure, by introducing the number of elements in $S_{\varepsilon}$ at a distance smaller than $L$ from the reference site, for some fixed $\varepsilon \ll 1$. This is our proxy for the number of resonances $1+\psi(L)$. Notice that 1 is added because there is at least one resonant site, even in the localized region, as the site is resonant with itself. It can be shown \cite{support_set} that $K_S(\varepsilon)$, and, hence, $\psi$ {\it in the delocalized phase} (generally multifractal), scales as $K_S\sim\psi\sim K^{D}=K_{0}^{L\,D}\gg 1$ where 
\begin{equation}\label{D_1}
 D=\frac{\partial\ln(1+\psi)}{\partial\ln K}\approx \frac{\partial\ln(\psi)}{\partial\ln K}. 
\end{equation}
The dimension $D$ is, in turn, easily determined numerically by the eigenfunctions Shannon entropy \cite{support_set}:
\begin{equation}\label{S_1}
D=D_{1}=\frac{dS}{d\ln K},\;\;\;\;S=S_{1}=-\Big\langle\sum_{r<L}\varphi^{2}(r)\,\ln \varphi^{2}(r)\Big\rangle
\end{equation}
Deeply in the {\it localized phase} the expected number of resonances within a distance $L$ decays exponentially $\psi\sim K_0^{-L/\xi}\sim K^{-\alpha}$ which gives $D=\frac{\partial\ln(1+\psi)}{\partial\ln K}\simeq\frac{\partial\psi}{\partial\ln K}\sim K^{-\alpha}$ as well. 

We note that in the limit $K\rightarrow\infty$,  $D$ is   $L$-independent and equal to zero in the localized phase, equal to 1 in the ergodic phase, and is a number $0<D<1$ in the multifractal phase. We now write an equation for the variable $\psi(L)$. The function $\psi(L)$ has to decrease exponentially in the localized region, namely when $\psi\ll 1$, and, in the delocalized region to be at most $K$. Our RG equations must have two fixed points; one at $\psi=0$ (localized phase) and another one at $\psi=K-1\approx K$ (delocalized phase). We write therefore our second equation as $\frac{d\psi}{d\ln L}=\psi\ln \psi \, \gamma(K,\psi)$,  where the function $\gamma(K,\psi)$ should obey the following property in the localized phase:
\begin{eqnarray}
   && \gamma(K,\psi)\to 1\quad   \mathrm{for\ all\ }\psi\ll 1.
\end{eqnarray}
This property ensures that in the localized region one can have arbitrary localization length: $\psi=K_0^{-L/\xi}$. In the delocalized phase $\gamma(K,\psi\gg1 )$ should obey the property $\gamma(K,K)=1$  in order to ensure the stable fixed point $\psi=K$.

Eliminating $L$  in favour of $K$ wth the help of Eq.(\ref{Eq-K}), we can write a single equation: 
\begin{equation}
    \frac{d\ln\psi}{d\ln K}=\frac{\ln \psi}{\ln K} \, \gamma(K,\psi).
    \label{eq:psikRG}
\end{equation}
  
In order to make further progress we need a form of the function $\gamma(K,\psi)$. As $\psi$ is not readily obtained from the numerics we instead use $D(L)$ defined in Eqs.~(\ref{D_1},~\ref{S_1}) as an implicit function of $K,\psi$. The whole idea of this paper is to follow the RG flow of $D(L)$ using the correspondence between RG flow and finite-size flow, and find the unknown RG functions from the numerics. We will show that even in the ergodic phase $D(L)$ is a non-trivial function that plays the same role as the dimensional conductance $g(L)$ does in the original `gang-of-four' work \cite{abrahams1979scaling}. 

The RG equation for this quantity defines the $\beta_{D}$ function
\begin{equation}
    \frac{d\ln D}{d\ln K}=\beta_D(D,K),
\label{eq:def_beta}    
\end{equation}
where $\beta_{D}(D,K)=(\gamma(K,\psi)-1)/\ln K$.
The function $\beta_D(D,K)$, unlike $\gamma(K,\psi)$, can be easily extracted from numerical data, and we make it now the main object of our study. First of all, notice that deep in the localized phase the scale-invariant law $D\sim K^{-\alpha}$ means that the localized phase is a line of fixed points at $D=0$ where $\beta=-\alpha$. The critical point $W=W_c$ corresponds to $\alpha=0$. This is the first of many similarities we will find with the Kosterlitz-Thouless phase transition with $\sqrt{D}$ being analogous to fugacity.

\begin{figure}[t]
    \centering
\includegraphics[width=0.49\textwidth]{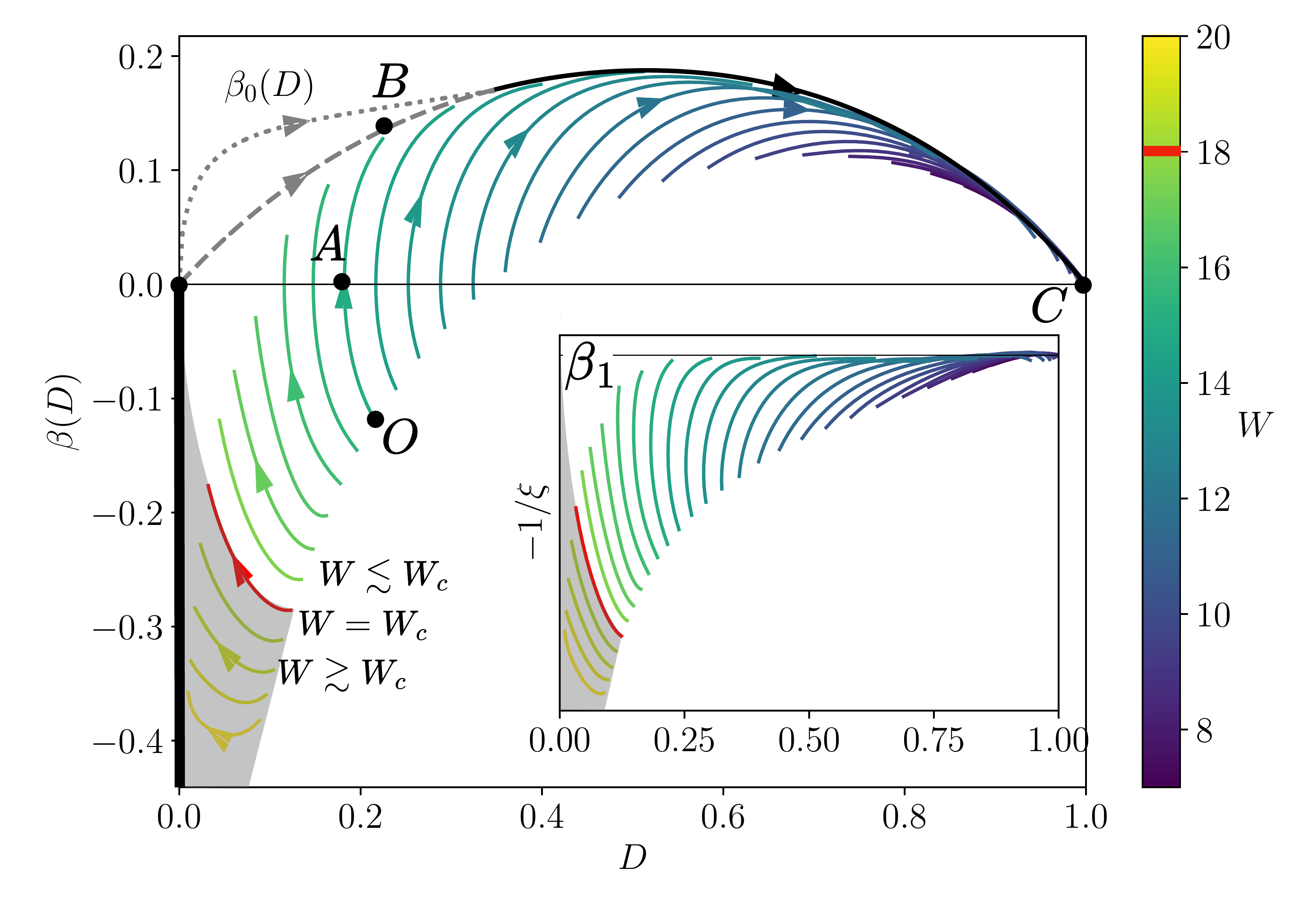}
    \caption{{\it (Main)} Numerical data for $\beta_D(D,K)$ (color corresponds to disorder, arrows indicate increasing system size $K$) and two possible behaviors of the function $\beta_0(D)$ (dashed and dotted lines). For values of $D\leq 0.3$, the curves are gray to emphasize that the shape is dependent on the fitting function used, {\it i.e.}\ either $\beta_0\propto D$ (dashed) or $\beta_0\propto \sqrt{D}$ (dotted). {\it (Inset)} The function $\beta_1(D,K)$. The same numerical results for the $r$ spectral parameter can be found in the Supplemental Material.   The existence of a 'single-parameter arc' $\beta_{0}(D)>0$ implies that the non-ergodic extended phase does not exist in the thermodynamic limit \cite{tikhonov2016Anderson,tikhonov2016fractality} but the multifractal behavior with $0<D<1$ is observed at finite sizes in the vicinity of the localization transition \cite{de2014anderson}. The critical exponent $\nu=1/2$ \cite{zirnbauer1986localization, mirlin1991localization,Tikhonov2019Critical} corresponds to evolution in the vicinity of point A in both scenarios. In scenario I we predict this exponent to be sub-leading at very large system sizes where the dominant behavior in the delocalized phase corresponds to $\nu=1$ \cite{kravtsov2018non,Pino2020Scaling}  }
    \label{fig:beta_01}
\end{figure}

Moving to $W<W_c$ we see from the numerical evidence (see Fig.~\ref{fig:beta_01}) that the curves tend to a single curve $\beta_0(D)$ and this allows us to make the central observation of the analysis presented in this paper, namely that the function $\beta_D$ can be divided into two, conceptually different pieces
\begin{equation}
    \beta_D(D,K)=\beta_0(D)+\beta_1(D,K),
\end{equation}
where
\begin{equation}
\label{eq:max}
    \beta_0(D)=\max_{K}\beta_D(D,K).
\end{equation}
By virtue of this definition $\beta_{1}(D,K)$ is negative.
In contrast, $\beta_0(D)$ does not depend on $K$ and {\it it is positive} $\beta_0(D)\geq 0$, vanishing for $D= 0$ and $D=1$. It can be extracted from the numerically obtained data (see Fig.~\ref{fig:beta_01}) by maximizing   $\beta_{D}(D,K)$  at fixed $D$, {\it i.e.} along vertical lines in Fig.~\ref{fig:beta_01}.
Let us stress that the different values of $\beta_D(D,K)$ at different $K$ correspond to different orbits and therefore to different $W$ (there is a one-to-one correspondence between $(D,K)$ and $(D,W)$).

A similar analysis on the (rescaled) $r$-parameter is shown in the Supplemental Material. In Fig. 6 of the Supplemental Material we also show that the function $\beta_0 (D)$ does not depend on the initial connectivity of the RRG, and thus {\it it defines unambiguously the universality class of the Anderson model in infinite dimensions.}

The emergence of a one-parameter scaling $\beta_{0}$ occurs because of large loops in the RRG: indeed if one considers a finite Cayley tree, the fractal dimension would saturate to a finite value between $0$ and $1$ (in the RG language, there is a line of fixed points $D\in[0,1]$)~\cite{tikhonov2016fractality}. In the RRG, the presence of loops favors the flow towards the ergodic fixed point $D=1$ along the "single-parameter arc" $\beta_{0}(D)$ which
 describes the true asymptotic limit of the RG flow and the reaching of the emergent one-parameter scaling regime:

\begin{equation} 
    \frac{d\ln D}{d\ln K}=\beta_0(D).
    \label{eq:one-param}
\end{equation}
The function $\beta_1$ describes the evolution of the system at the beginning of the flow, before large loops are encountered, and both parameters, $D$ and $K$, are necessary to describe the scaling in the critical region and in the localized regime.  
 
The function $\beta_0(D)$ has two zeros, at the fixed points $D=0,1$. A fit of the numerical data with a simple polynomial (see Supplemental Material for details) which vanishes in 0 and 1, we get $\beta_0(D)= 1-D+O((1-D)^2)$ for $D\to 1$. The derivative at $D=1$ is 1 within the statistical error (see Fig.~\ref{fig:beta_01}, solid black curve). In fact, in the one-parameter scaling theory of Ref.~\cite{abrahams1979scaling}, $D$ is an analytic function of $g$. One can prove close to $d=2$, that $D=1-1/g+O(\epsilon)+O(g^{-2})$ \cite{altshuler2024renormalization}, but we believe this expansion to be valid for all $d$ and even in the limit $d\to\infty$.\footnote{The vanishing of the $1/g$ term can possibly occur for a different symmetry class of the Hamiltonian.} So, since $D\simeq 1-1/g+...$ for $g\to\infty$, and $g\sim L^{d-2}$, in $d$ dimensions this slope should be $(d-2)/d$~\cite{altshuler2024renormalization}. In the limit $d\to\infty$ we recover the observed slope 1. 

The situation near $D\to 0$, is more complicated. Our numerical data allow a reliable extraction of $\beta_0$ only down to $D\simeq 0.3$ so we must guess the form of $\beta_0$ down to $D=0$, where it must vanish. The two simplest situations are either a simple zero $\beta_0(D)\propto D$ or $\beta_0\sim D^{1/2}$. We will see later that these two possibilities imply two different physical pictures.

Other functional forms of $\beta_0$ close to $D=0$ are also possible within our theory, but the main point of this paper stands: there exists {\it a function} $\beta_0(D)$, which describes the one parameter scaling flow of $D$ away from the critical point $D=0$ and towards the ergodic critical point $D=1$, thus excluding the multifractal behavior in the thermodynamic limit. This function must be calculable from first principles, but not necessarily from a Cayley tree calculation. In fact, on the Cayley tree the fractal dimension $D_1$ can take any value in $[0,1]$~\cite{monthus2011anderson,tikhonov2016fractality} which is possible if $\beta_0(D)=0$ and the flow is generated by $\beta_1(D,K)$. We believe the function $\beta_0(D)$ has not appeared in previous literature on the Anderson model on the RRG, although its analog $\beta(g)$, with $g$ being dimensionless conductance, is central in the discussion of finite dimensional systems~\cite{abrahams1979scaling}.

The one-parameter scaling motion is the solution of Eq.~\eqref{eq:one-param}, obtained by integrating the differential equation by separation of variables. The result for the two different {\it ansatzes} is shown in Fig. 5 of the Supplemental Material. 

Notice that as long as the evolution is on the one-parameter segment and Eq.~\eqref{eq:one-param} holds one obtains
\begin{equation}
\ln(K/K_{in})=\int_{D_{in}}^{D}\frac{dD'}{D'\,\beta_{0}(D')}\equiv\ln F(D)-\ln F(D_{in}),
\end{equation}
where $F$ results from the integration and $\ln K_{in}$ is a length scale for a system to evolve during a two-parameter regime from the initial condition at small system size through the minimum of $D$ and subsequently during a single-parameter regime to $D_{in}\lesssim 1$. Now, inverting the function $F(D)$, we obtain:
\begin{equation}\label{volumen}
D(K)=F^{-1}(K/K_{c}),
\end{equation}  
where $K_{c}=K_{in}/F(D_{in})$ is a critical volume. Eq.~\eqref{volumen} corresponds to a {\it volumic} scaling  \cite{Mata2017Scaling}, which therefore holds in the delocalized phase as soon as evolution proceeds along the single-parameter arc.
 
The function $\beta_1(D,K)$ is dominant, and it has a simple form near the critical line $W_c \simeq 18.17$ \cite{Tikhonov2019Critical,parisi2019anderson}, but it becomes negligible for sufficiently large system sizes far from the critical point (inset of Fig.~\ref{fig:beta_01}) yielding to the one parameter regime. 

\section*{The critical region}
In the delocalized region, the critical behavior which describes the divergence of $K_B$ when $W\to W_c$, is obtained by looking at the ``time" $\ln(K/K(L=L_{0}))$ it takes one to reach the fixed point $D=1-\epsilon$ with any given accuracy $\epsilon\ll 1$. The approach to the fixed point $D=1$ happens in two steps: first, the motion is governed by $\beta_1$, since $|\beta_1| \gg \beta_0$. Then, after the orbit approaches the asymptotic curve $\beta_0(D)$ at some scale $K\sim K_B$, the motion is described by Eq.~\eqref{eq:one-param}. 
Referring to the main panel of Fig.~\ref{fig:beta_01}, we have two times to sum: the first one is the time necessary to go from the initial condition $K_0, D(K_0)$ (or equivalently $K_0,W$) until the one parameter curve $\beta_0(D)$, corresponding to the path $OA + AB$ in the figure. Then one moves along the $\beta_0$ curve till reaching the delocalized fixed point $D=1$, corresponding to the path $BC$ in the figure. The times along both $OA$ and $AB$ diverge algebraically when $W\to W_c$ with the same exponent $\nu=1/2$  independently of the behavior of $\beta_{0}(D)$ near the origin, but the corresponding exponents for motion along $\beta_{0}(D)$ in the vicinity of the point $B$ are different for different choices of the function $\beta_{0}(D)$ at $D\ll 1$.

Along the branch OA, when $W\to W_c$ from the delocalized region, we must expect according to Ref.~\cite{tikhonov2016Anderson, Altshuler2016Nonergodic} $\beta_D(D,K) = 0$ for some value $D_A$ which corresponds to the minimum in the dependence $D(L)$ at a fixed $W$.
As can be seen in the right panel of Fig. 8 in the Supplemental Material, $D_A$ as a function of $W$ is almost linear throughout the entire range of $W$ where the minimum is observed. In fact, a simple linear extrapolation of the fit gives $W_c=18.0$, a good estimate of the Anderson transition point. More specifically, the fit for $W\to W_c$ reads
\begin{equation}
    D_A=\eta \, \frac{W_c-W}{W_c}, \quad\mathrm{with}\quad \eta=1.1\pm 0.1,
    \label{eq:D0num}
\end{equation}
a particularly simple result, which seems to hold {\it mutatis mutandis} for higher connectivities as well. 
 
The system spends a large amount of RG time near the minimum $D_A$, diverging when $D_A\to 0$. In order to enter the truly ergodic, one-parameter region, one needs to have system sizes $K\gg K_A$. After that initial slow-down, the fractal dimension starts moving towards its final value $D=1$.

In order to give a quantitative dependence of $K_A$ on $D_A$ we need a model of the function $\beta_1(D,K)$. The clue to finding this model is to remember that, at $D=0$, $\beta$ defines the localization length. In fact, $D\propto K^{\beta(D=0)}=K_0^{-L/\xi}$. So, $\beta=-1/\xi$, and it must approach a constant at $K\rightarrow\infty$. Thus $d\beta/d\ln K\rightarrow 0$ in this limit. However, in this very limit also $D\rightarrow 0$ (see also Fig.~\ref{fig:beta_01}). Therefore we come to a conclusion that 
$d\beta/d\ln K = \phi(D)$, where $\phi(D)$ must vanish at $D=0$.  
In principle, any function $\phi(D)$ like $D^{\alpha}$ $(\alpha>0)$ may do the job. However, it is the simplest choice $\phi(D)=c\,D$ with $c\approx 1$ that corresponds to the numerics (see Fig. 10 in Supplemental Material).
 
This leads to the two equations:
\begin{eqnarray}
    \frac{d\ln D}{d\ln K}=\beta,\\
    \frac{d\beta}{d\ln K}=c D,
\end{eqnarray}
which are solved implicitly by
\begin{equation}
\label{eq:D-DA}
    \frac{1}{2}\beta^2(D,K)=c\,(D-D_{A}).
\end{equation}

The time to pass the region around $D=D_A$ (from the region $D\gg D_A$) can be found from the solution of the equation:
\begin{equation}
    \frac{d\ln D}{d\ln K}=-\sqrt{D-D_A},
\end{equation}
which is 
\begin{equation}
    D(\ln K)=\frac{D_A}{\cos^2\left(\frac{1}{2}\sqrt{D_A}\ln(K_A/K)\right)},
\label{eq:D_turn}
\end{equation}
Notice that the presence of the square root guarantees that the integral $\int dD/(D\,\beta(D))$ is convergent at $D=D_{A}$, and therefore  $D_A$ is {\it not a fixed point} although the RG time spent in its vicinity diverges when $D_A\to 0.$ In fact, $D=D_{A}$ is a turning point where the function $D(\ln K)$ reaches the minimum. Thus the solution Eq.~\eqref{eq:D_turn} can be extended from $K<K_A$ to $K>K_A$, where $\beta>0$.

Now we can compute the RG time $\ln(K_A/K(L=L_{O}))$ to go from O to A, where $D_{O}=D(K(L=L_{O}))\sim O(1)$ while $D_A\to 0$. This means that the argument of $\cos^{2}$ should be close to $\pi/2$ in order to compensate for the smallness of the numerator. We find, therefore, from Eq.~\eqref{eq:D_turn}: $\ln(K_{A}/K(L=L_{O}))=\pi D_A^{-1/2}\sim \pi (1-W/W_c)^{-1/2}$, which gives 
\begin{equation}
K_A\sim e^{\frac{\pi}{\sqrt{1-W/W_c}}}. 
\end{equation}
This divergent volume corresponds to a critical exponent $\nu=1/2$. The RG time to pass from $A$ to $B$ has the same type of divergence. 

Notice that Eq.~\eqref{eq:D_turn} can be written also in terms of $L=\ln K$ and  $x=\sqrt{D_A}\,L=\sqrt{1-W/W_{c}}\,L$, for $D_A\to 0$ as
\begin{equation}\label{lin-sc}
    D(L)=D_{c}(L) \frac{(\pi x/2)^{2}}{\sin^{2}(\frac{\pi x}{2})}\quad (0<x<1),
\end{equation}
where $D_{c}(L)$ is the fractal dimension on the critical line:
\begin{equation}
    D_c(L)=\frac{1}{(\ln K)^2}.
    \label{eq:DcL}
\end{equation}
Eq.~\eqref{lin-sc} has a canonical form of  the {\it linear} (i.e.\ $L$ as opposed to the volumic $K$ \cite{Mata2017Scaling}) scaling associated to the critical region. 
Summarizing our discussion on the type of finite-size scaling, one can claim that the volumic scaling, Eq.~\eqref{volumen}, corresponds to the single-parameter scaling governed by $\beta_{0}(D)$ in the extended phase, while the linear scaling is a signature of an essentially two-parameter RG in the localized and the near-critical extended regime.

The critical dependence of $D$ on $K$ in Eq.~\eqref{eq:DcL} is a universal law: it was observed for the first time in Ref.~\cite{sierant2023universality} for several ensembles of expander graphs of constant or even random connectivity, and it can be seen in Fig. 9 of the Supplemental Material.  Let us mention that, in the context of RG analysis with scaling variable $K$, the behavior on the critical line corresponds to a {\it marginally irrelevant variable} since the inverse logarithmic dependence on $K$ corresponds to a critical exponent $y=0^{-}$.

\begin{figure}[t]
    \centering
    \includegraphics[width=0.49\textwidth]{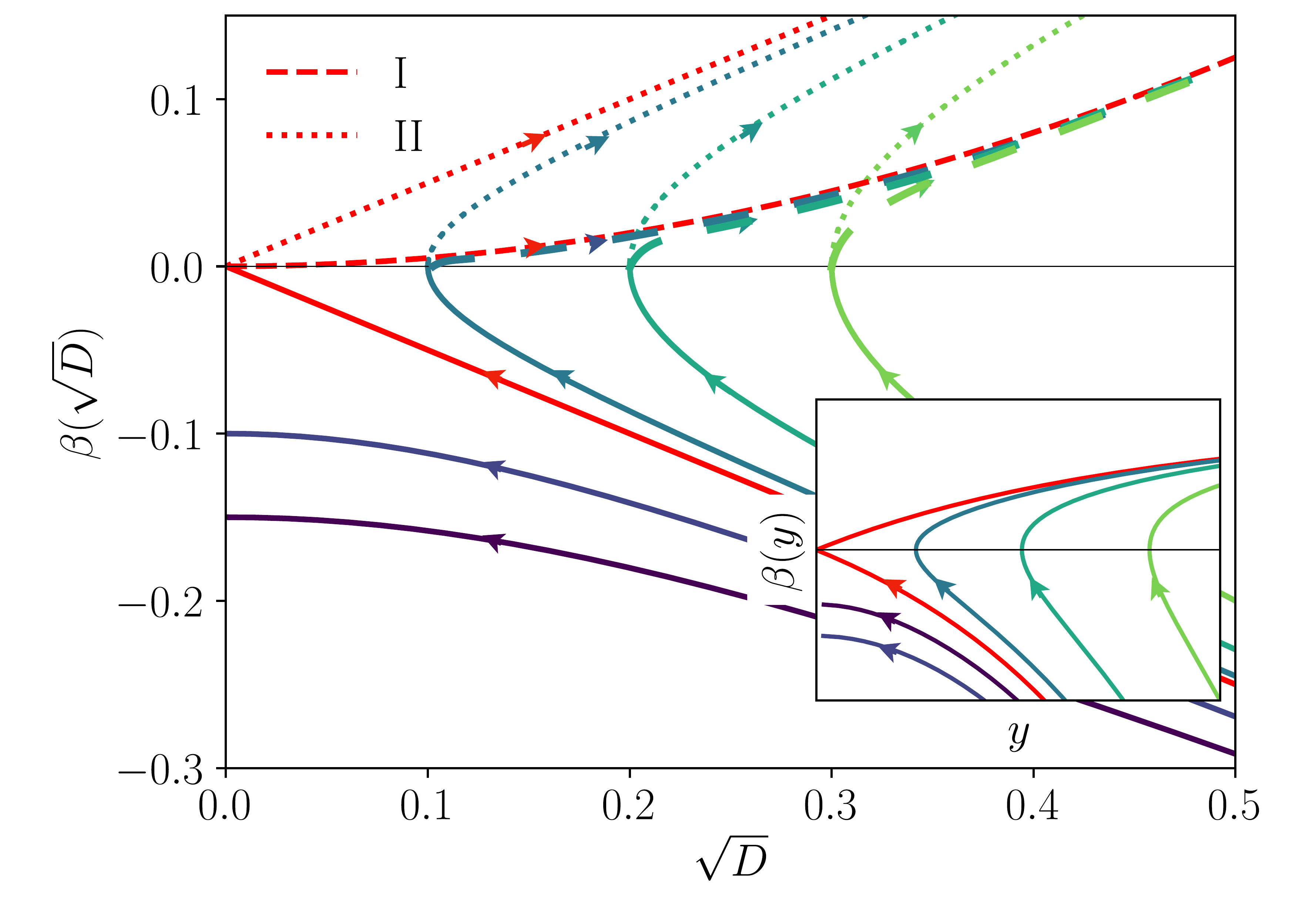}
    \caption{ The $\beta$-function $\beta(\sqrt{D})=d\ln(\sqrt{D})/d\ln K$ versus $\sqrt{D}$ in the vicinity of the fixed point $D_{c}=0$ and the Kosterlitz-Thouless flow for the fugacity $y$, shown in the inset. The two possible forms of the single-parameter asymptotic behavior $\beta_{0}(\sqrt{D})$ are shown by the red dashed and dotted lines. Scenario II gives the same behavior for $\beta(\sqrt{D})$ as for the $\beta$-function $d\ln(y)/d\ln L$ of the fugacity in the Kosterlitz-Thouless RG, while scenario I gives a different behavior and a new exponent emerges. Notice the existence of a minimum of both $D$ and $y$ and a square root behavior of the $\beta$-functions near this minimum. Such a behavior is not possible in the single-parameter scaling where the $\beta$-function must be smooth and single-valued. Notice also that the localized phase is similar to a superfluid one of Kosterlitz-Thouless RG, with the line of fixed points $D=0$, $\beta(D)<0$ and $y=0$, $\beta(y)<0$ respectively. Both of them cannot be obtained within the single-parameter scaling with $D_{c}=0$ or $y_{c}=0$. Thus the second parameter  $K$ is relevant for RRG, in contrast to parameters with irrelevant exponents for the Anderson localization problem on $d$-dimensional lattice which only decorates the RG flow in the localized phase at small sample sizes
    \cite{altshuler2024renormalization}.
    }
    \label{fig:beta_KT}
\end{figure}

If these were the only divergent timescales in the motion from the point O to the final region close to $D=1$ the critical exponent would be $\nu=1/2$  ~\cite{zirnbauer1986localization,mirlin1991localization}. 
However, we now encounter two possibilities depending on the behavior of the function $\beta_0(D)$ close to $D=0$, which we present here (see also Fig.~\ref{fig:beta_KT}).

1) {\it First scenario:} $\beta_0(D)\propto D$. In this case, the motion from $A$ to $B$ (see Fig.~\ref{fig:beta_01}) intercepts $\beta_0$ at $D\sim D_A$. The motion from $B$ to any $D=O(1)$ takes time $ \sim D_A^{-1}\sim (1-W/W_c)^{-1}$ in view of Eq.~\eqref{eq:one-param}. Notice that in this scenario there are two critical lengths $L_{1}$ and $L_{2}$: one ($L_{1}$) with the exponent $1/2$ for reaching the single-parameter scaling   and the other one ($L_{2}$) with the exponent $1$, such that for $L_{1}<L<L_{2}$ the system behaves as approximately multifractal one with $D_{1}\approx D_{A}$. This behavior is reminiscent of the one predicted for the Rosenzweig-Porter random matrix ensemble associated with RRG \cite{LN-RP-RRG21}. 

Two critical lengths in the localization problem on RRG with the exponents 1/2 and 1 were reported recently in Ref.~\cite{Mata2020two}. However, this work is concerned with the localized phase on RRG, rather than the extended phase that we study.

2) {\it Second scenario:} $\beta_0(D)\propto D^{1/2}$. In this scenario, $\beta_0$ is the analytic continuation of the critical $\beta_1(D,K)$. No new length is introduced and therefore $\nu=1/2$, which is what one obtains by solving for the fixed point of iterative, mean field equations in Refs.~\cite{tikhonov2016Anderson,tikhonov2021AndersonMBL,zirnbauer1986localization,mirlin1991localization}.

We notice that the finite-size scaling exponent $\nu$ is determined by the behavior of $\beta_{0}(D)$ near the fixed point $D_{c}=0$. This behavior is very sensitive to boundary conditions and for example it changes drastically between a finite Cayley tree and an RRG. In fact, it is known that in the finite Cayley tree the whole delocalized phase is multifractal 
\cite{monthus2011anderson,tikhonov2016fractality}. In our language this means $\beta_{0}=0$ and the two-parameter scaling dominates the whole delocalized region. We believe then that the mean-field approach is too rough to distinguish between the two scenarios that we propose. On top of that, an identification is made between the exponent controlling the behavior of the diffusion coefficient \cite{zirnbauer1986localization} (and the typical local density of states \cite{Tikhonov2019Critical}) for the {\it infinite} system and the exponent $\nu$ of the {\it finite-size} scaling \cite{Mata2020two, Pino2020Scaling}, and the exponent that controls the localization radius in the localized phase. All of them describe different physics and they do not have to be the same.

The behavior $\beta_0(D)\sim \pm\sqrt{D}$ around $D=0$ is also reminiscent of the Kosterlitz-Thouless flow for the square of fugacity $y^{2}$~\cite{1972JETP...34..610B,1971JETP...32..493B,1973JPhC....6.1181K}. This is seen easily by passing from $D$ to $\sqrt{D}$ as we do in Fig.~\ref{fig:beta_KT}. Notice also the similarity of RG flow in the localized phase to that in the superfluid phase of the Kosterlitz-Thouless flow, the negative semi-axis $D=0, \beta(D)<0$ ($y=0, \beta(y)<0$) being the line of fixed points in both cases.

Figure \ref{fig:beta_01} unequivocally shows that current numerical results cannot yet rule out any of these scenarios, and further analytical and numerical work is needed to resolve the issue. Recent analytical developments~\cite{arenz2023wegner,zirnbauer2023wegner}, in which a new field theory of localization was proposed and studied beyond the weak-coupling regime, give a picture that in many ways resembles the one presented in this paper. Further work is needed in this direction.

\section*{Conclusions}
We have presented 
 a real space renormalization group analysis of the Anderson model on infinite dimensional graphs in which one of the parameters is a finite-size fractal dimension $D(L)$ and the other, the running connectivity $K$.  In particular, the fact that the critical point has all the properties of the localized phase makes the RG flow close to it quite peculiar, distinguishing it from the RG flow for finite dimensions $d$ described for the first time in Ref.~\cite{abrahams1979scaling} and from a typical Wilson-Fisher fixed point. We have introduced the division of the flow into a function that is responsible for the one-parameter scaling $\beta_0(D)$ and which describes the approach to the ergodic fixed point $D=1$, and a two-parameter scaling part $\beta_1(D,K)$ which describes the remaining motion, in particular close to the minimum of $D(K)$ and to the critical point $D=0$. We believe our work provides the correct perspective to look at Hamiltonians with localized critical points, among which it is believed there are many models displaying many-body localization phenomenology \cite{Deluca13}, showing that the whole beta function needs to be considered, and not only its linearization close to the fixed point. We also provided a clean way to describe non-perturbative effects in such systems, which go beyond the tree-geometry results. Among directions for future work, let us mention finally the possibility of doing a controlled $1/d$ expansion of the Anderson model around the mean-field $d=\infty$ result discussed in this paper.

\begin{acknowledgments}

\emph{Acknowledgments.~---}
The authors are deeply grateful to Anton Kutlin for many interesting discussions and collaborations on closely related topics and to Piotr Sierant for sharing the numerical data that have been used in this work and for fruitful discussions.
C.V.\ is also grateful to Anushya Chandran, Sarang Gopalakrishnan, David A. Huse, and David M. Long for stimulating discussions and collaboration on related topics and to Boston University and Princeton University for their kind hospitality during the work on this project. A.S.\ acknowledges financial support from PNRR MUR project PE0000023-NQSTI. V.E.K.\ is grateful to Ivan Khaymovich for fruitful discussions and support from Google Quantum Research Award ``Ergodicity breaking in Quantum Many-Body Systems". V.E.K. is grateful to KITP, University of Santa Barbara for hospitality. This research was supported in part by the National Science Foundation under Grant No. NSF PHY-1748958. 
 
\end{acknowledgments}

\bibliography{references.bib}

\pagebreak
\widetext
\newpage

\section{Supplemental material - Renormalization Group Analysis of the Anderson Model on Random Regular Graphs}

\subsection*{Data analysis for the fractal dimension}

As mentioned in the main text, the numerical data for the fractal dimension are extracted from the participation entropy, defined as
\begin{equation}
    S_q = \Big\langle \frac{1}{1-q}  \log_2 \sum_{i=1}^{K} |\varphi(i)|^{2q} \Big\rangle,
\end{equation}
and in particular, in this work, we used $S_q$ for $q\to 1$, which is the von Neumann entropy
\begin{equation}
\label{eq:vonNeumann}
    S_1 = - \Big \langle \sum_{i=1}^{K}  |\varphi(i)|^{2} \log_2 |\varphi(i)|^{2} \Big\rangle.
\end{equation}
From $S_1$, the fractal dimension can be extracted as $D(L)=d S_1 /dL$ (or equivalently as the $S_1(L) = D(L)L + c(L)$~\cite{sierant2023universality}). We have absorbed here the $\ln K_0 = \ln 2$ factor in the definition of $S_q$.
Having at our disposal finite increments in the system size $L$, we computed the fractal dimension as
\begin{equation}
    D(L) = S_1(L+1)-S_1(L)
\end{equation}
and we then consider, for the numerical analysis, $\overline{D}(L+1/2) = (D(L+1)+D(L))/2$.

In order to obtain continuous curves, we \emph{interpolated} the numerical values of $\overline{D}(\ln K)$ with two different fits, depending on the value of $W$. Denoting $\ln K=x$ for brevity, we use a Pad\'e-like function for the fit at $W\leq 17$  
\begin{equation}
    f(x) = \frac{x^3 + c_1 x^2 + c_2 x + c_3}{x^3 + d_1 x^2 + d_2 x + d_3},
\end{equation}
so that $f(x)=1$ for $x \to \infty$, as it should in the delocalized phase. For larger values of $W$ instead, we use a fourth-order polynomial fit in $1/x$, that perfectly fits the numerical data. We use these functions to compute the $\beta$-function using the definition and to produce the plot in Fig.~2 of the Letter, employing only the range of system sizes for which we have numerical data so that we are just interpolating, without extrapolations.

\begin{figure}
    \centering
    \includegraphics[width=0.6\textwidth]{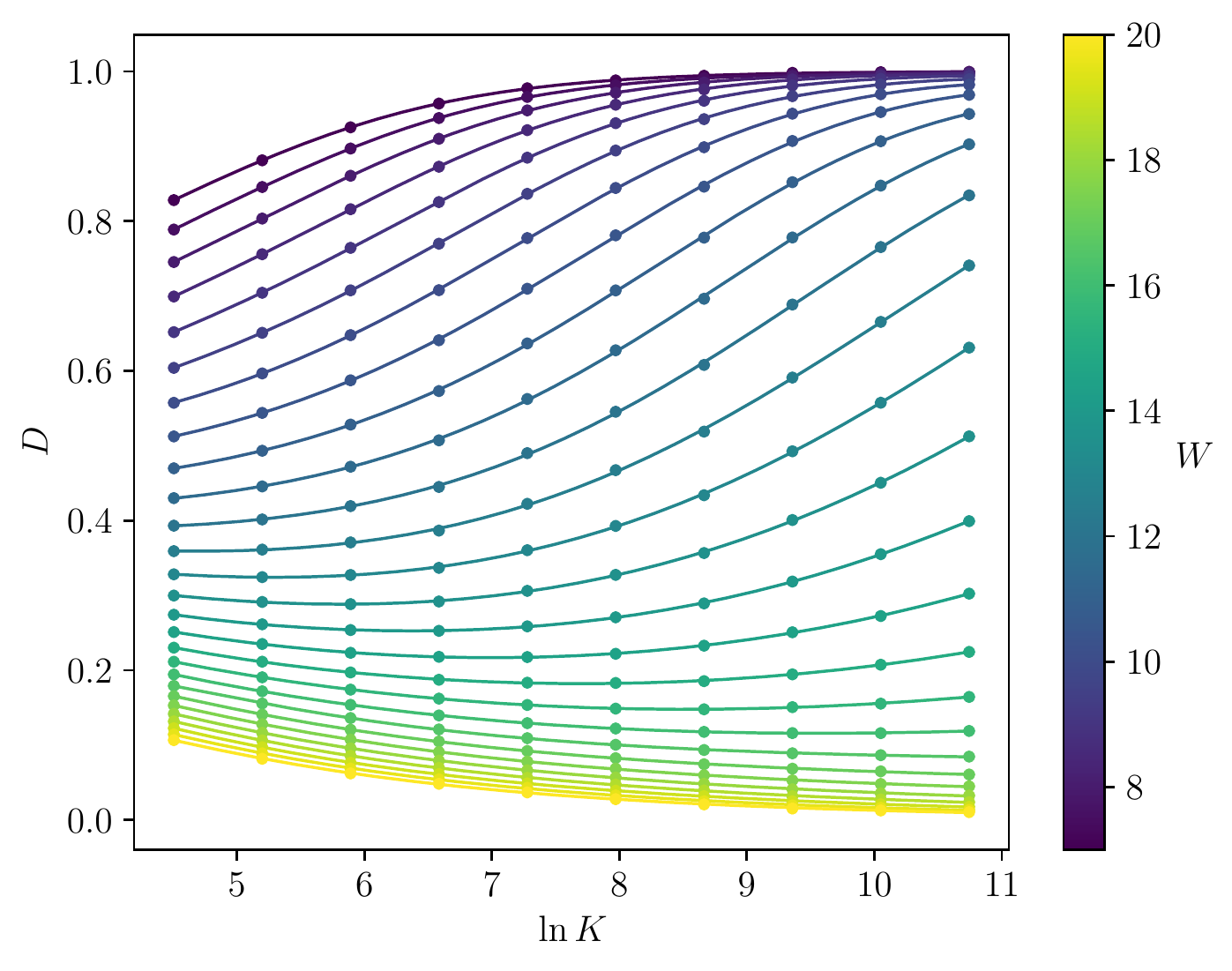}
    \caption{Numerical data of the fractal dimension as a function of $K$ (dots) and their interpolation using Eq.~\eqref{eq:fit_g} and Eq.~\eqref{eq:fit_h}.}
    \label{fig:D_logK}
\end{figure}

We then used these data also to determine two possible forms for the function $\beta_0$. The function $\beta_0$ has to fit the envelope that the numerical data are generating for $D\gtrsim 0.3$, and we numerically do so by fitting the set of points that are obtained by considering, for any small interval $dD$ ($D \geq 0.3$), the maximum value of $\beta(D,K)$ for all $W$. We use two different fitting functions, having different behaviors in $D=0$. The dashed line in Fig.~\ref{fig:beta_01} is obtained through
\begin{equation}
\label{eq:fit_g}
    g(x) = a_1 \, x(1-x) + a_2 \, x(1-x)^2 + a_3 \, x(1-x)^3 + a_4 \, x^2(1-x) + a_5 \, x^2(1-x)^2 + a_6 \, x^3(1-x),
\end{equation}
while the dotted line is obtained using the fitting function
\begin{equation}
\label{eq:fit_h}
    h(x) = b_1 \, \sqrt{x}(1-x) + b_2 \, \sqrt{x}(1-x)^2 + b_3 \, x(1-x) + b_4 \, x^{3/2}(1-x).
\end{equation}
We report the fitting coefficients in Tab.~\ref{tab:fit}, and the resulting interpolations compared with the bare data in Fig.~\ref{fig:D_logK}

\begin{table}[ht!]
\begin{center}
\begin{tabular}{||c c | c c||} 

 \hline
 $g(x)$ &  & $h(x)$ & \\ [0.5ex] 
 \hline\hline
 $a_1$ \qquad & 0.153 & $b_1$ \qquad & 0.986 \\ 
 \hline
 $a_2$ \qquad & 0.235 & $b_2$ \qquad & 0.266 \\
 \hline
 $a_3$ \qquad & 0.476 & $b_3$ \qquad & 2.517 \\
 \hline
 $a_4$ \qquad & 0.285 & $b_4$ \qquad & 2.429 \\
 \hline
 $a_5$ \qquad & 0.366 &  &  \\
 \hline
 $a_6$ \qquad & 0.501 &  &  \\ [1ex] 
 \hline
\end{tabular}
\end{center}
\caption{Fit coefficients for the functions $g(x)$ (Eq.~\eqref{eq:fit_g}) and $h(x)$ (Eq.~\eqref{eq:fit_h})}
\label{tab:fit}
\end{table}

Let us remark that the function $\beta_0$ obtained using the fitting function $g(x)$ in Eq.~\eqref{eq:fit_g} turns out to have the symmetry $D \to 1-D$ within the precision of the fit even if $g(x)$ doesn't have such symmetry.

\subsection*{$\beta$-function for the $r$-parameter}

The same analysis performed on the fractal dimension can be reproduced for the $r$-parameter, once rescaled so that it ranges between $0$ and $1$ as
\begin{equation}
    \phi = \frac{r-r_P}{r_{WD} - r_P},
\end{equation}
where $r_P \simeq 0.386$ and $r_{WD} \simeq 0.5307$. The same procedure outlined above for the fractal dimension $D$ is applied to the data for the $r$-parameter, and the resulting $\beta$-function is displayed in the left panel of Fig.~\ref{fig:beta_r}.

Let us mention that near $\phi=1$ the envelope of the functions $\beta(\phi)$ is different from $\beta_0(D)$ and, in particular, a best fit is obtained assuming that the derivative $\beta'(\phi\to 1)$ diverges logarithmically. This gives a qualitative approach $\phi\to 1$
\begin{equation}
    \phi\simeq 1-e^{-(K/K_0)^{a}},
\end{equation}
with $a=0.43\simeq 1/2$. Such difference is still explained in the one-parameter scaling, and it originates from the non-linear dependence of $\phi$ on $D$ (or vice-versa) near $D,\phi =1$, when plotted together as in Fig.~\ref{fig:D_phi}. We leave a more detailed analysis of these features for future work.

We report in Fig.~\ref{fig:r_0} the values of $\phi_A$ such that $\beta(\phi_A) = 0$, as we did for the fractal dimension. Also in this case a linear fit gives a good prediction for the critical value of the disorder. Moreover, the critical behavior of the $\beta_1(\phi)\sim -\sqrt{\phi}$ implies that
\begin{equation}
    \phi_c(K)\propto\frac{1}{(\ln K)^2},
\end{equation}
which has been already observed in \cite{sierant2023universality}. We report in Fig.~\ref{fig:collage_piotr} the plot presented in~\cite{sierant2023universality}, showing the $1/L^2$ approach to $r_P$ of the $r$-parameter, which turns out to be universal for many models of random graphs.

\begin{figure}
    \centering
    \includegraphics[width=0.6\textwidth]{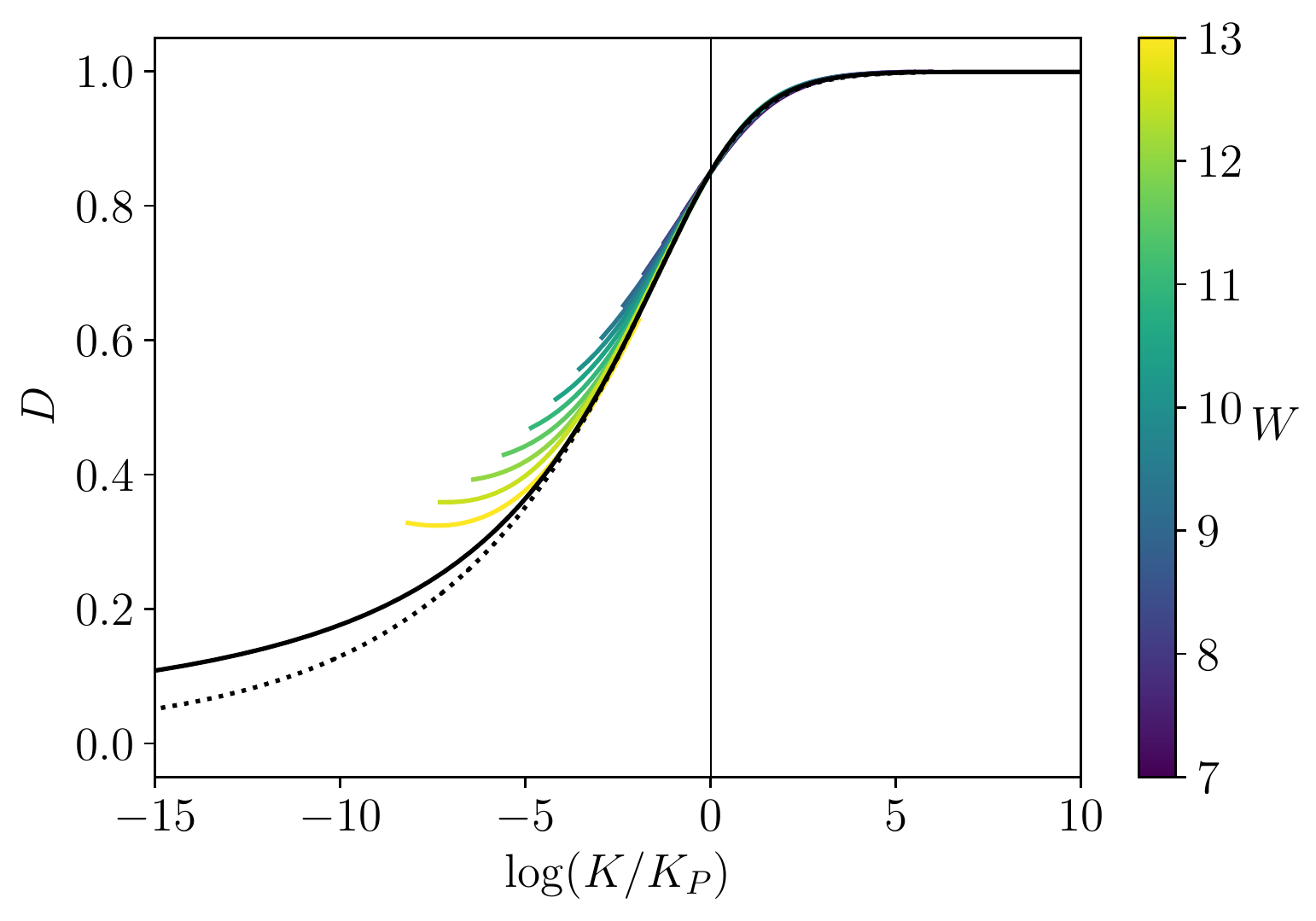}
    \caption{The function $D(\ln K)$ in the one-parameter scaling regime, for two best fits for $\beta_0$, the one vanishing linearly in $D=0$ (solid) and the one with square root singularity at the origin (dotted). The integration constant $K_P$ is chosen in such a way that all the curves intersect in one point $K=K_P$, where $D(K_P)=0.85$ for all curves, numerical and analytical. The numerical data are shown as thin colored lines for $W\in [7,13]$.}
    \label{fig:DK_asymp}
\end{figure}

\begin{figure}
    \centering
    \includegraphics[width=0.49\textwidth]{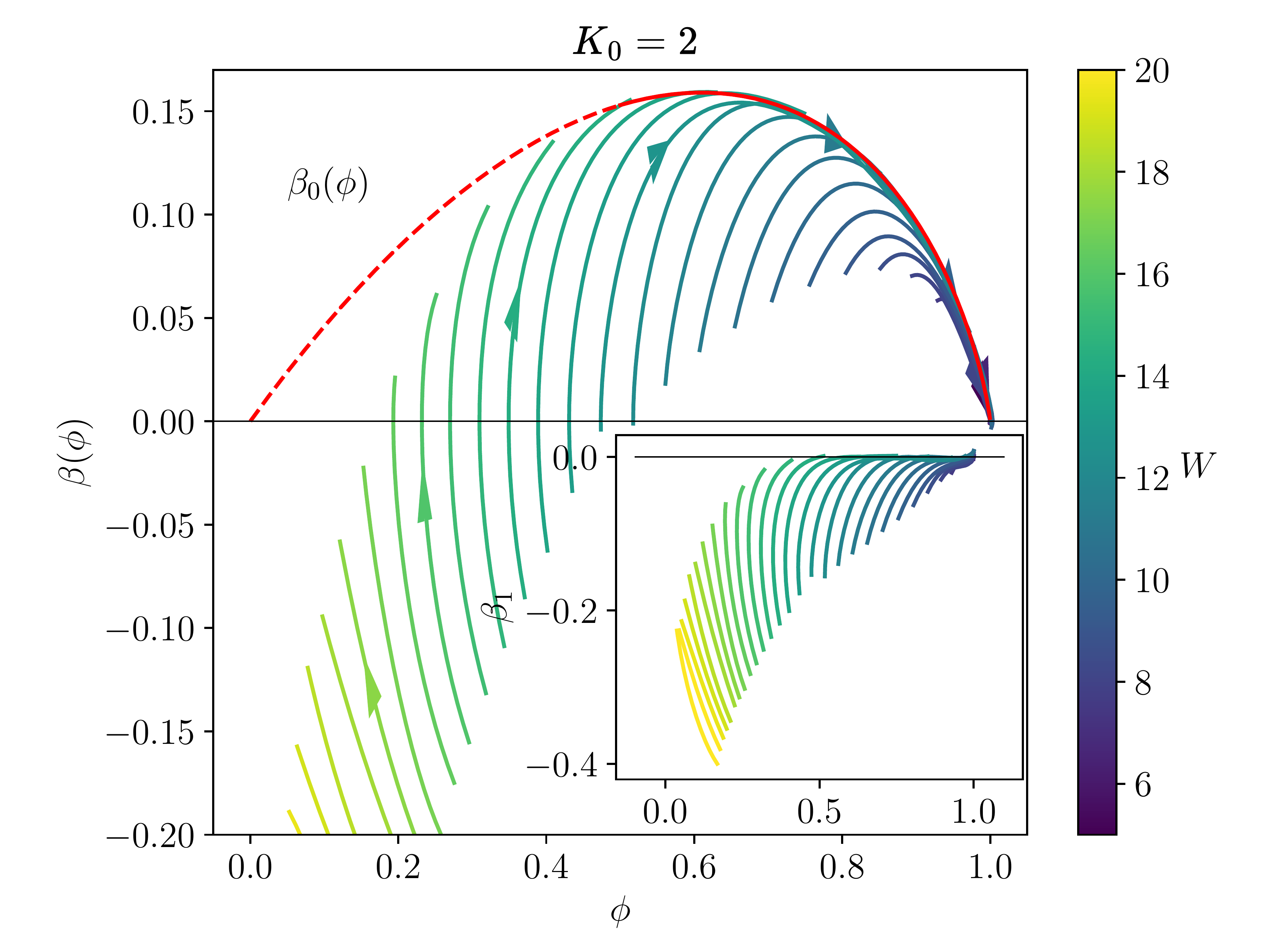}
    \includegraphics[width=0.49\textwidth]{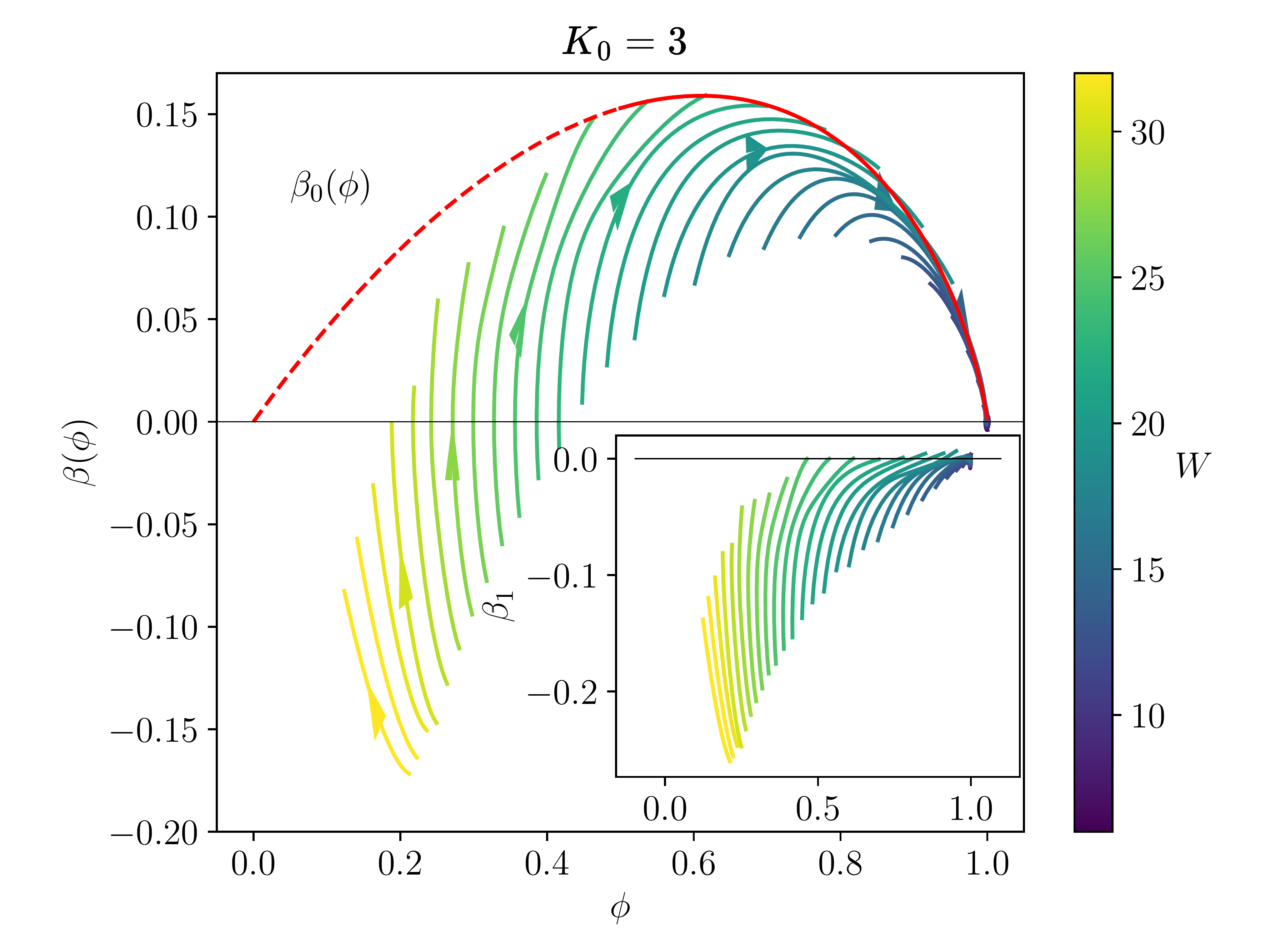}
    \caption{{\it (Left)} $\beta$-function for $\phi$, i.e. the rescaled $r$-parameter. The data analysis performed is the same that has been done for the fractal dimension. The red curve is obtained by fitting the envelope with the function $z(x) = -a (1-x)\ln(1-x) + b x (1-x) + c x^2 (1-x)$ ($a\simeq0.43$, $b\simeq0.06$, $c\simeq -0.11$). In particular, the $\beta$-function approaches $\phi=1$ with an infinite derivative, as it is also confirmed by the right panel. For $\phi<0.5$ we just extrapolated the fitting function used for the points at $\phi>0.5$, and therefore can be not accurate given the limits of the numerics.
    {\it (Right)} Same analysis performed for a RRG with $\mathcal{D}=4$. The $\beta_0(D)$ function is the same as in the $\mathcal{D}=3$ case, and it is in perfect agreement with the data, supporting the universality of the function $\beta_0$.}
    \label{fig:beta_r}
\end{figure}

\begin{figure}
    \centering
    \includegraphics[width=0.49\textwidth]{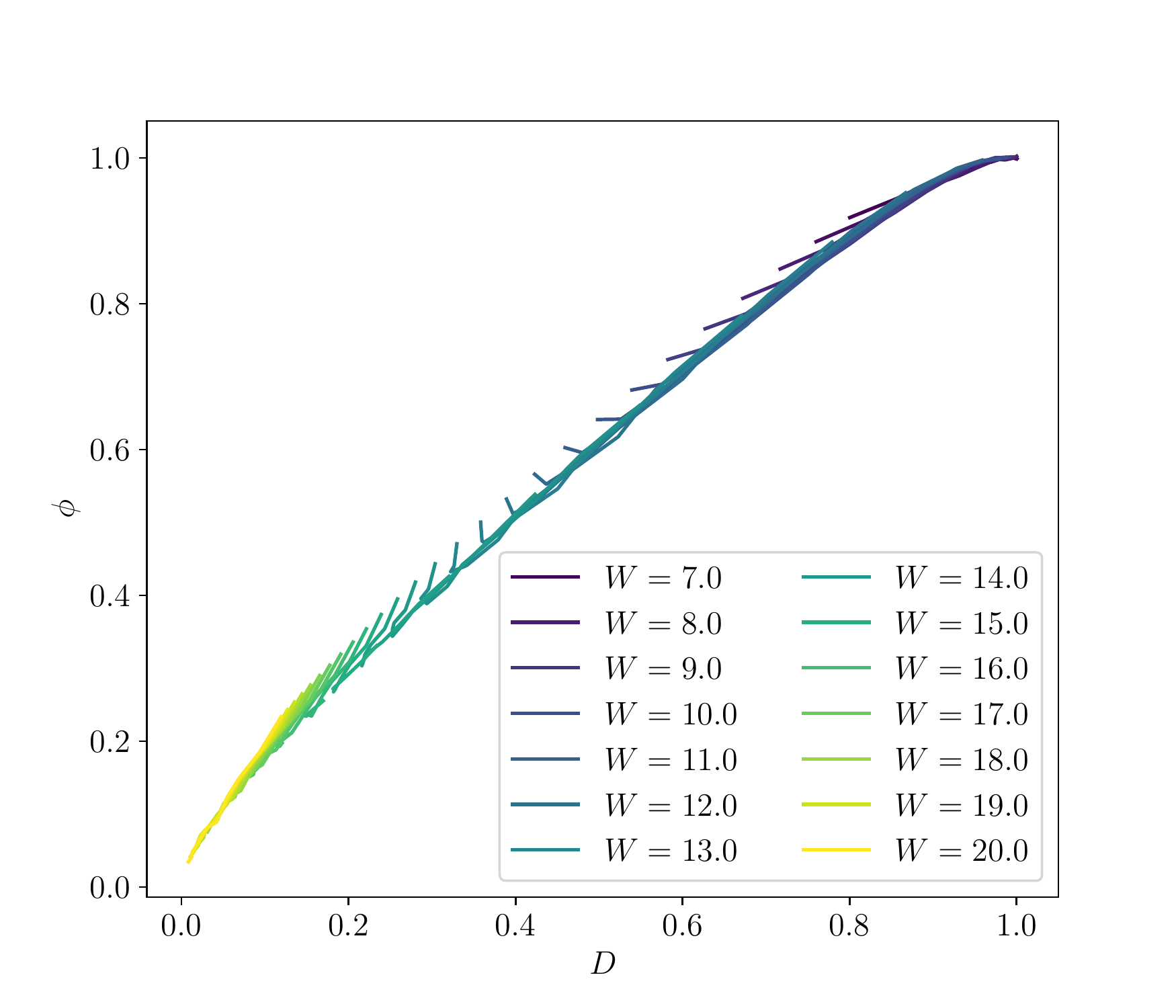}
    \caption{$\phi$ vs $D$. For intermediate values of $D$ (and $\phi$), the two quantities are almost perfectly proportional one to the other. Near $D,\phi \sim 1$ however, the curve has a vanishing derivative.}
    \label{fig:D_phi}
\end{figure}

\begin{figure}
    \centering
    \includegraphics[width=0.49\textwidth]{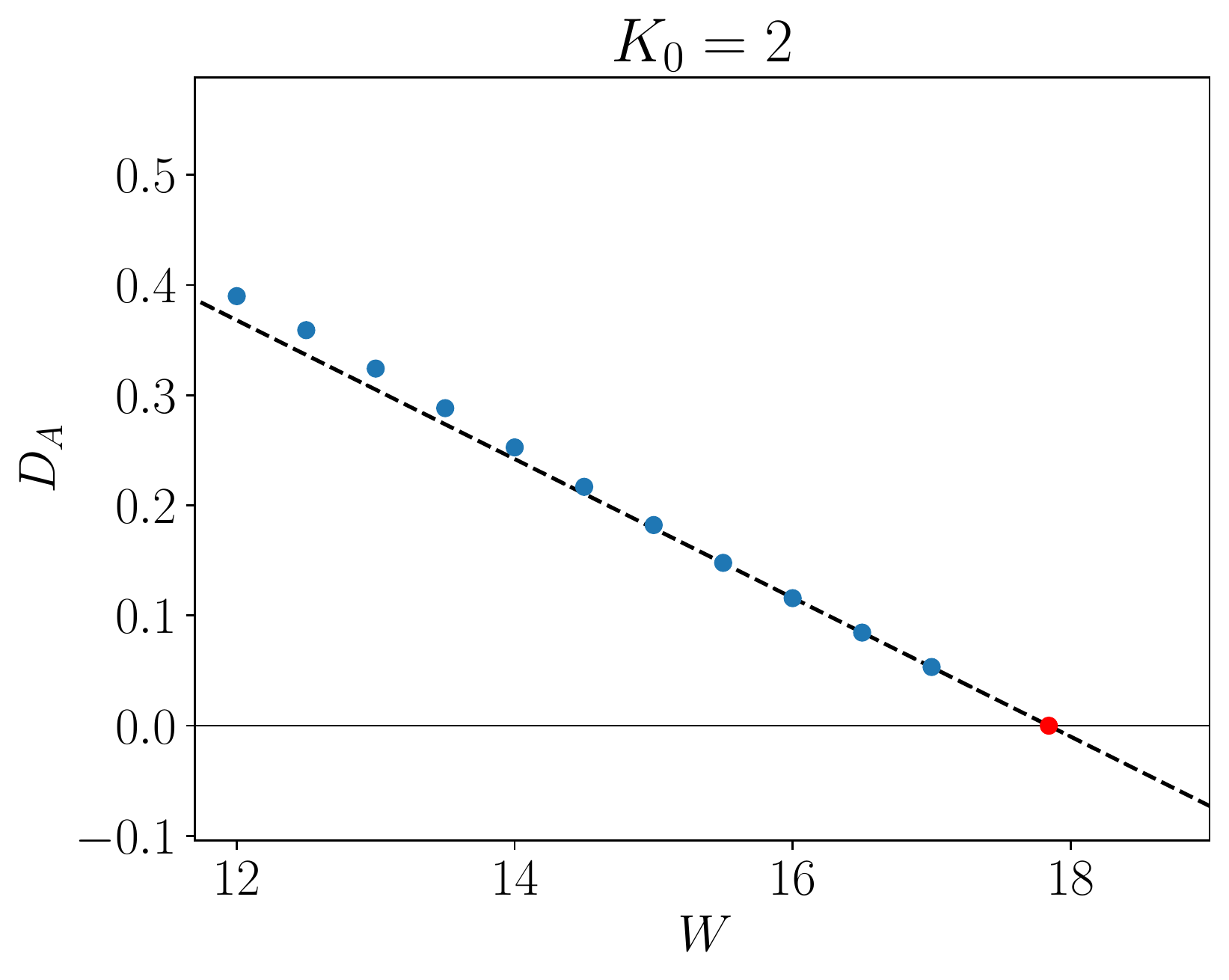}
    \includegraphics[width=0.49\textwidth]{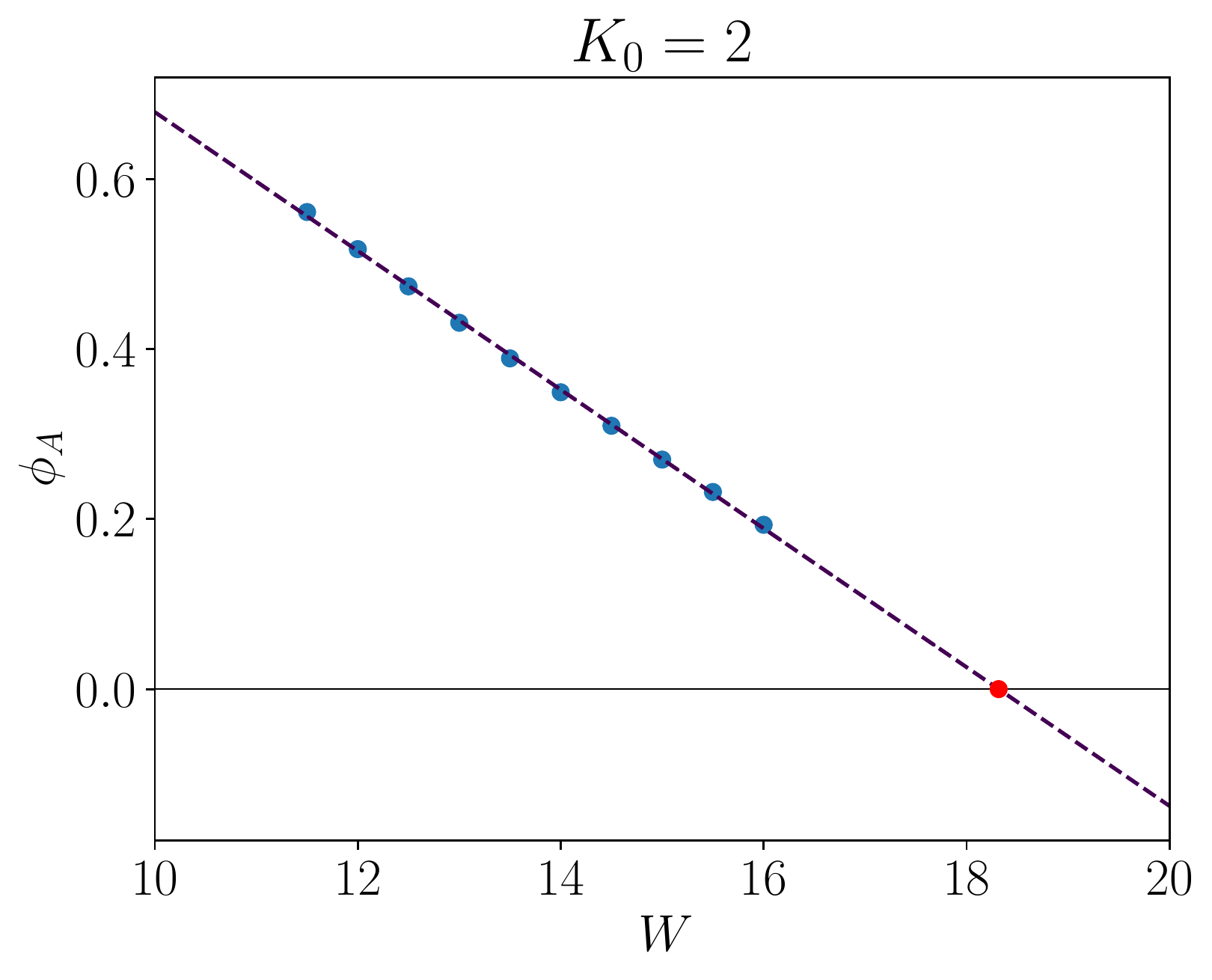}
    \caption{Values of $\phi_A$, i.e. the values of the rescaled $r$-parameter such that $\beta(\phi_A)=0$, for different values of $W$ (blue dots). A linear extrapolation gives a critical value for the disorder $W_c = 18.3 \pm 0.1$ (red dot), which is in very good agreement with the known position of the transition $W_c=18.17$.}
    \label{fig:r_0}
\end{figure}

\begin{figure}
    \centering
    \includegraphics[width=0.8\textwidth]{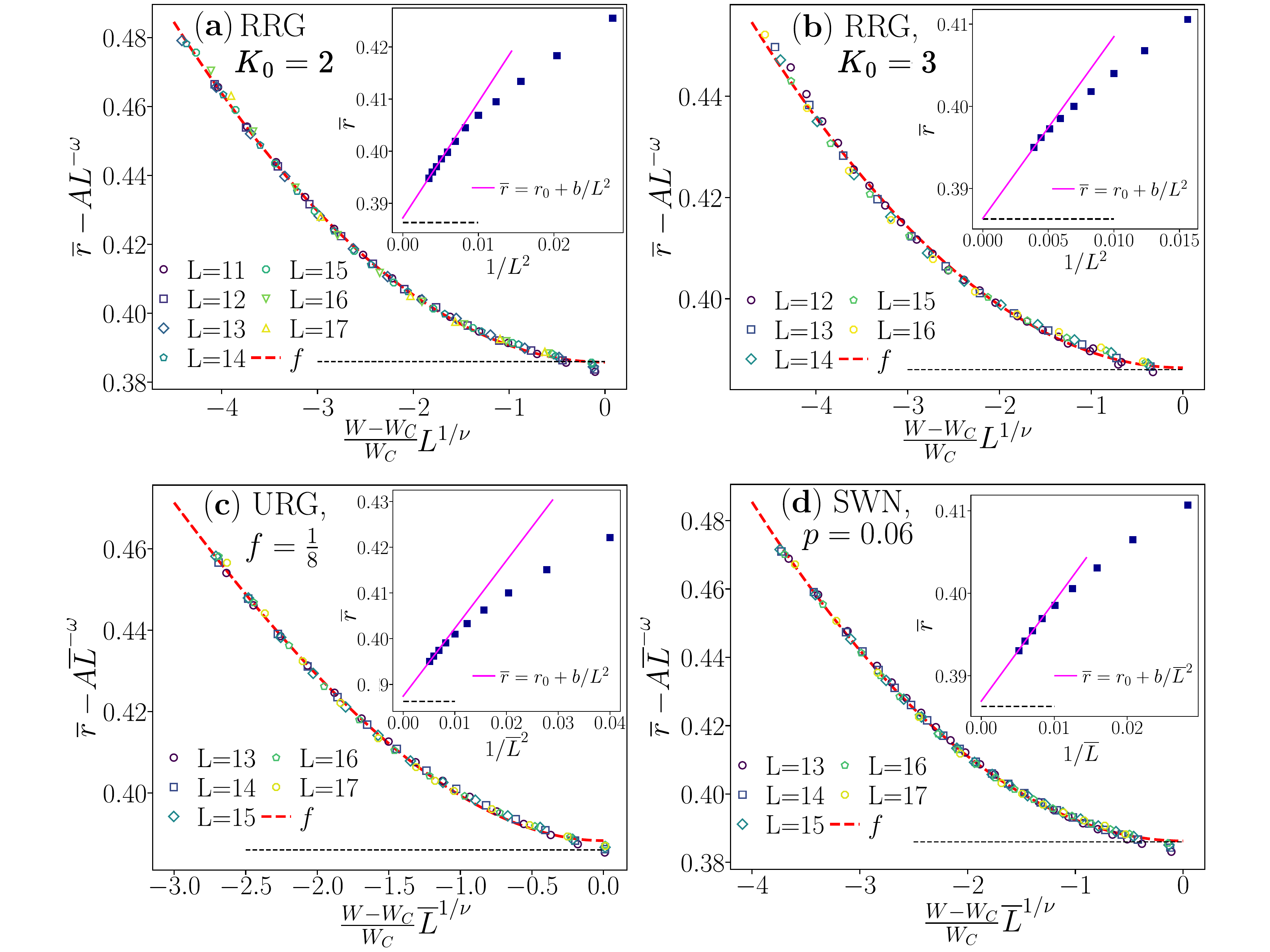}
    \caption{Figure taken from Ref.~\cite{sierant2023universality}. The main figures show the collapse of the $r$-parameter for different values of $W$ and system size. The collapse is obtained by setting $\omega=2$ and $\nu=1$ (see~\cite{sierant2023universality} for a description of the critical exponents and data collapse). The insets show the behavior of the average gap ratio at the critical point, displaying a $1/L^2$ scaling, which is predicted by our renormalization group equations. Different subfigures ((a), (b), (c), and (d)) correspond to different types of network, respectively RRGs with $K_0=2$, $K_0=3$, uniform random networks and small-world networks (notice that here $D$ in the plots is the vertex coordination number $K_0+1$).}
    \label{fig:collage_piotr}
\end{figure}

\section{$\beta$-function in the critical region}

\begin{figure}
    \centering
    \includegraphics[width=0.8\textwidth]{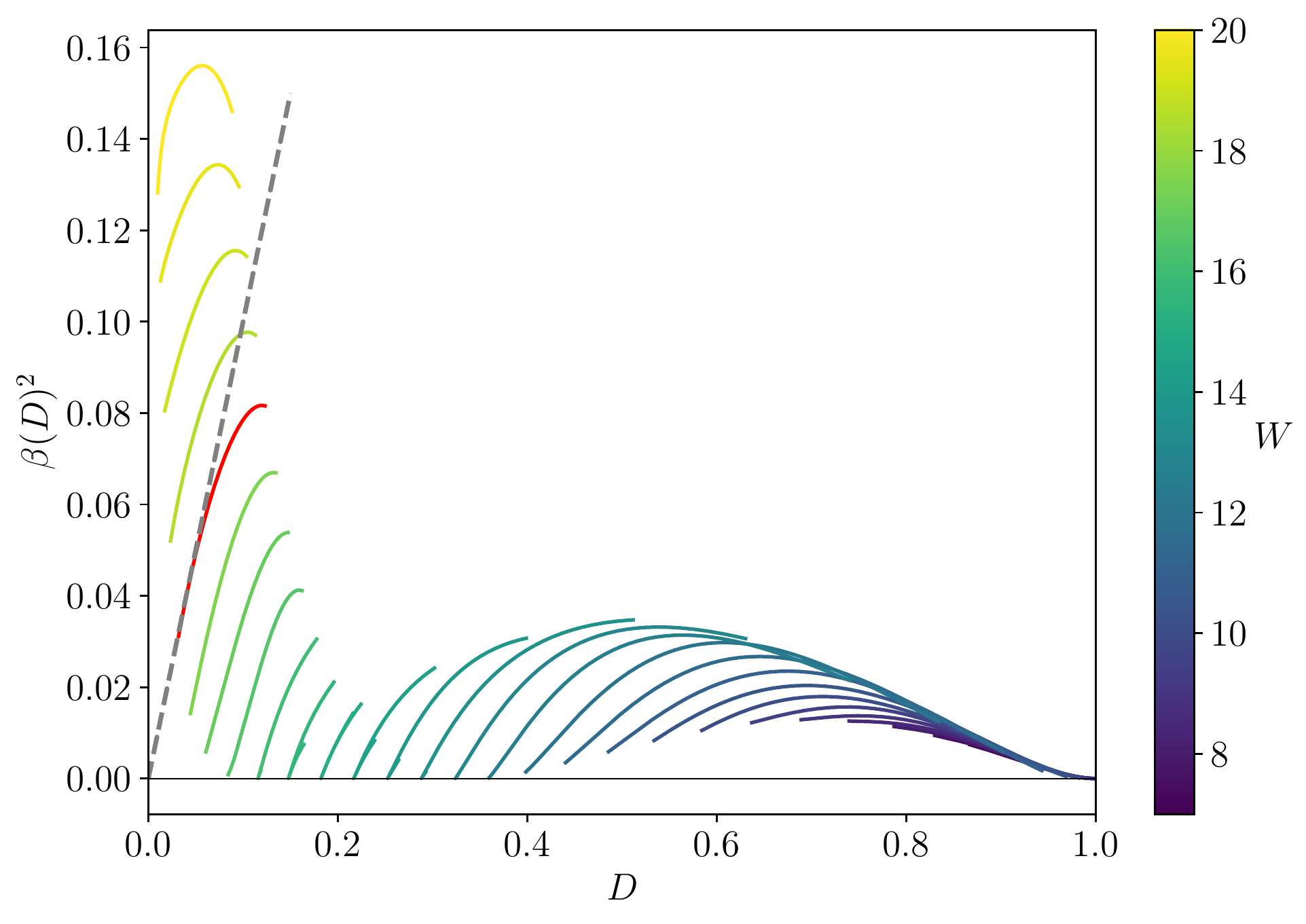}
    \caption{Plot of $\beta^2(D)$ vs $D$. As explained in the text, we notice that the slope of the curves near $D=D_A$ (i.e. $\beta \simeq 0$) does not depend significantly on $D_A$ (or equivalently $W$). Moreover, the critical curve, represented in red, follows the curve $\beta^2(D) = D$ (dashed line in the plot), meaning that $c=1/2$.}
    \label{fig:beta2}
\end{figure}

In this Section, we want to elaborate on Equation (16) for the $\beta$-function presented in the main text. In particular, we want to show that, for small $D$, in Eq. (15) the constant $c$ is of order unity and does not depend on $D$ and/or $D_A$. In fact, in general, we could have
\begin{equation}
    \frac{d \beta}{d \ln K} = c(D/D_A) D.
\end{equation}
We are interested in the behavior of $c(D/D_A \simeq 1)$, which means that we can trade $D$ with $D_A$. In Fig.~\ref{fig:beta2} we show the numerical data for $\beta^2(D)$. If $c(1)$ had a dependence on the initial condition, say $c(1) \simeq D_A$, one would have that the slope of $\beta(D=D_A)$ would depend on $D_A$, and for our example we would have that the slope vanishes at the critical point, where $D_A = 0$. If instead $c(1) = O(1)$ for all $D_A$, the slope would not depend on $W$ and it would be finite at the critical point. We can see in Fig.~\ref{fig:beta2} that this is actually the case, with, in particular $c=1/2$, proving that $\beta_1(D,K) = \pm \sqrt{D-D_A} + O(D-D_A)$. Let us mention, however, that different expressions for $c(1)$ are possible for different models, for example the Cayley tree.

\end{document}